\documentclass[preprint,prc,aps,showpacs,showkeys,groupedaddress,floatfix]{revtex4-1}
\usepackage{epsfig}
\usepackage{dcolumn}
\usepackage{bm}
\usepackage[latin5]{inputenc}
\usepackage{graphics}
\usepackage{graphicx}
\usepackage{amssymb}
\usepackage{amsmath}
\usepackage[english]{babel}

\begin{document}
\title{Surface Interaction Effects to a Klein-Gordon Particle
Embedded in a Woods-Saxon Potential Well in terms of Thermodynamic Functions}
\author{B.C. L\"{u}tf\"{u}o\u{g}lu}
\affiliation{Department of Physics, Akdeniz University, 07058
Antalya, Turkey}
\date{\today}
\begin{abstract}
Recently, it has been investigated how the thermodynamic functions vary when the surface interactions are taken into account for a nucleon which is confined in a Woods-Saxon potential well, with a non-relativistic point of view. In this manuscript, the same problem is handled with a relativistic point of view. More precisely, the Klein-Gordon equation is solved in presence of mixed scalar-vector generalized symmetric Woods-Saxon potential energy that is coupled to momentum and mass. Employing the continuity conditions  the bound state energy spectra of an arbitrarily parameterized well are derived. It is observed that, when a term representing the surface effect is taken into account, the character of Helmholtz free energy and entropy versus temperature are modified in a similar fashion as this inclusion is done in  the non-relativistic regime. Whereas it is found that this inclusion leads to different characters to internal energy and specific heat functions for relativistic and non-relativistic regimes.
\end{abstract}
\maketitle

\section{Introduction}\label{sec:Intro}

In the nineteenth century, the properties of a macro system had been examined with a macroscopic point of view, so-called classical thermodynamics, in terms of measurable thermodynamic variables like temperature, pressure, volume etc.. On the other hand, the development of quantum theory in the twentieth century caused to reconsider the macro systems from the microscopic point of view. That approach, is known as statistical mechanics, is made up of Hamiltonian dynamics and mathematical statistics. In the latter methodology, partition function plays the central role and the dynamics of a physical system is described with a set of quantum states.  Moreover, the thermal properties of the system are determined by the multiplicity of these quantum states within partition function that depends on temperature. Helmholtz free energy can be introduced with partition function. It is well known that Helmholtz free energy can be introduced with partition function and its various derivatives refer to several thermodynamic quantities, i.e. internal energy which is the mean energy of the system, entropy which is the measure of the disorder of the system and the specific heat which is the fluctuations in energy \cite{reifbook, santrabook}. As a consequence to determine thermodynamic quantities of the investigated system, let scientists gain a better understanding of nature. Thus in the last decade, an increasing interest is seen in this field. Note that  not only non-relativistic problems \cite{dong2006, vincze2015, ikhdair2013421}  but also relativistic ones are examined by various scientist \cite{ikhdair2013421, ikot2016131, suparmi2016, onyeaju2017132, hassanabadi2015, pacheco2003, pacheco2014, boumali2014, boumali2015, boumali2015dkp, larkin2015, arda2015}. Moreover, experimental studies was initiated to support the theoretical studies \cite{villafane2013}.

Among the potential energies that was subjected to be investigated in relativistic or non-relativistic equations, the Woods-Saxon (WS) potential differs from the others because it possesses an essential role in microscopic physics. Although it was originally proposed to calculate differential cross section of protons elastically scattered from medium or heavy nuclei \cite{woodsSaxon1954}, after having success of being consistent with experimental results, it was widely employed to the problems in the different fields of physics as nuclear   \cite{woodsSaxon1954, zaichenkoOlkhovskii1976, pereyPerey1968, michelNazarewiczPloszajczakBennaceur2002, michelNazarewiczPloszajczak2004, esbensenDavids2000, coban2012, ikhdair2013, gautam2014, brandanSatchler1997, soylu2015, koyuncu2017, salamon2016}, atom-molecule \cite{brandanSatchler1997, satchler1991} with relativistic \cite{kennedy2002, panellaBiondini2010, aydogduArda2012, guoSheng2005, guoZheng2002, rojasVillalba2005, hassanabadiMaghsoodi2013, yazarlooMehraban2016, chargui2016, salamon2017} or  non-relativistic \cite{pahlavaniAlavi2012, costaPrudente1999, fluge1994, saha2011, feizi2011, niknamRajabi2016} approaches.

When WS potential energy is used to describe the confinement of a nucleon in nuclei, the surface effects are not taken into account. In the vicinity of the surface, attractive or repulsive forces can be dominant to keep the nucleon inside the well around the surface or core. Such surface effects are being confronted in other physical processes, too. Such that, when a particle approaches to a nucleus, the incident particle should be suffered by surface forces before the core effects. Detailed discussions can be found in the twelfth chapter of Satchler    \cite{satchlerbook}.  The sum of volume and surface terms is called as the generalization of the WS potential energy. Generalized Woods-Saxon (GWS) potential energy has been subjected on many articles \cite{ lutfuoglu201617, lutfuoglu201721, lutfuoglu2018-1, lutfuoglu2018-2, candemirBayrak2014, bayrakAciksoz2015, bayrakSahin2015, lutfuogluAkdeniz2016, liendoCastro2016, berkdemirBerkdemir2005, badalovAhmado2009, gonulKoksal2007, kouraYamada2000, capakPetrellis2015, capakGonul2016, ikotAkpan2012, ikhdairFalayeHamzavi2013, Kobos1982, Boztosun2002, Boztosun2005, Kocak2010, Dapo2012}.

Recently, the thermodynamic functions in the presence of surface interaction terms are analyzed with a non-relativistic point of view \cite{lutfuoglu201617, lutfuoglu201721}. Basically, the solution of the Schr\"{o}dinger equation is used to obtain the energy spectra, which was provided in \cite{lutfuogluAkdeniz2016}.
In \cite{lutfuoglu201617} nucleon was confined in a light nucleus while in \cite{lutfuoglu201721} in a heavy one. Note that in both studies, bound state solution is examined for GWS potential well. Then, the calculated bound states energy spectra are divided to tight-bound and quasi-bound spectra. It is shown that, in heavy or light nuclei the effect of excluding and including the quasi-bound energy spectra vary the thermodynamic functions similarly.

In a very recent paper  \cite{lutfuoglu2018-1}, the thermodynamic functions of an alpha particle that is confined in a heavy nucleus are investigated by using the WS and GWS potential energies in the non-relativistic point of view. More precisely, the bound state solutions of Schr\"{o}dinger equation are calculated separately for both potential wells and the comparison of  the Helmholtz free energy, entropy, internal energy and specific heat functions are discussed in details within the aspect of statical mechanics. Note that this comparison is done non-relativistically and an investigation within a relativistic point of view is an open question. Answering this question is going to fill the gap in the literature. Hence, this is the main motivation for this study.

The present study is going to contribute to the field since its comparative results are very interesting and differ from the non-relativistic study. The first astonishing result is the wave numbers obey different constraints in relativistic and non-relativistic case. On the other hand, the number of existing microstates decrease similar to the non-relativistic problem. Thus Helmholtz free energy and Entropy functions vary similarly in relativistic case. The other thermodynamic functions, namely internal energy and specific heat functions, do not alter similarly. In relativistic case, the internal energy of the confined particle in the WS well is greater than the internal energy of the confined particle in GWS well. Note that in non-relativistic case it was opposite.  As expected the specific heat function has the contrary behavior since it is proportional to the change in the internal energy.

The manuscript is organized as follows.
Section \ref{sec:themodel}  starts with the Klein-Gordon(KG) equation with mixed scalar and vectorial potentials. Then, the one dimensional KG equation is solved in strong regime with generalized symmetric Woods-Saxon (GSWS) potential energy and the required conditions to confine the particle to any GSWS potential well are discussed.  In section \ref{sec:thermosection}, a brief definition of the thermodynamic functions that will be compared is given. In section \ref{sec:application}, an arbitrary potential well is constituted to calculate the energy spectrum for a confined neutral Kaon.   Meanwhile, the calculated energy spectra of neutral Kaon particle for the two potential energies are presented. Afterwards, the partition function is constructed and the plots of the resulting thermodynamic functions of the system are obtained. Finally, the effects of the surface interactions terms are discussed. In the conclusion section, section \ref{sec:conclusion}, it is mentioned that the relativistic approach results differ from non-relativistic one in terms of behaviors of internal energy and specific heat functions while Helmholtz free energy and entropy functions do not.

\section{The Model}\label{sec:themodel}
It is assumed that the KG equation is not only coupled with a vector potential that has a non-zero time component and vanishing spatial components, to the mass parameter by a scalar potential. In strong regime, if the vector and scalar potential energies have positively proportional magnitude,  the time-independent KG equation of a spinless particle with a mass $m$ and an energy $E$ can be written by \cite{greinerbook}
\begin{eqnarray}
  \Bigg[\frac{d^2 }{d x^2}+\frac{1}{\hbar^2c^2}\Big[ (E^2-m_0^2c^4)-2(E+m_0c^2)V(x)\Big]\Bigg]\phi(x) &=& 0.
\end{eqnarray}
The GSWS potential has been proposed with extra terms to the customary WS potential by
\begin{eqnarray}\label{gws}
  V(x)&=&\theta{(-x)}\Bigg[-\frac{V_0}{1+e^{-\alpha(x+L)}}+\frac{W e^{-\alpha(x+L)}}{\big(1+e^{-\alpha(x+L)}\big)^2}\Bigg] \nonumber \\
  &+& \theta{(x)}\Bigg[-\frac{V_0}{1+e^{\alpha(x-L)}}+\frac{W e^{\alpha(x-L)}}{\big(1+e^{\alpha(x-L)}\big)^2}\Bigg], \label{gsws}
 \end{eqnarray}
where $\theta{(\pm x)}$ are the Heaviside step functions \cite{lutfuogluAkdeniz2016}. The GSWS potential well is defined by four parameters, $\alpha$ determines the capability of the diffusion, where $L$ is the range of the effective forces.  The customary $V_0$  parameter is responsible for  the depth of the potential well besides it is proportional to measure of the surface interaction barrier $W$ by a dimensionless multiplier. The freedom of assigning positive or negative values to the multiplier, it is possible to study well with a barrier or pocket. In this study only  the positive case, where bound states can be categorized by tightly-bound and/or quasi-bound states, is investigated.

\begin{figure}[!htb]
\centering
\includegraphics[totalheight=0.5\textheight]{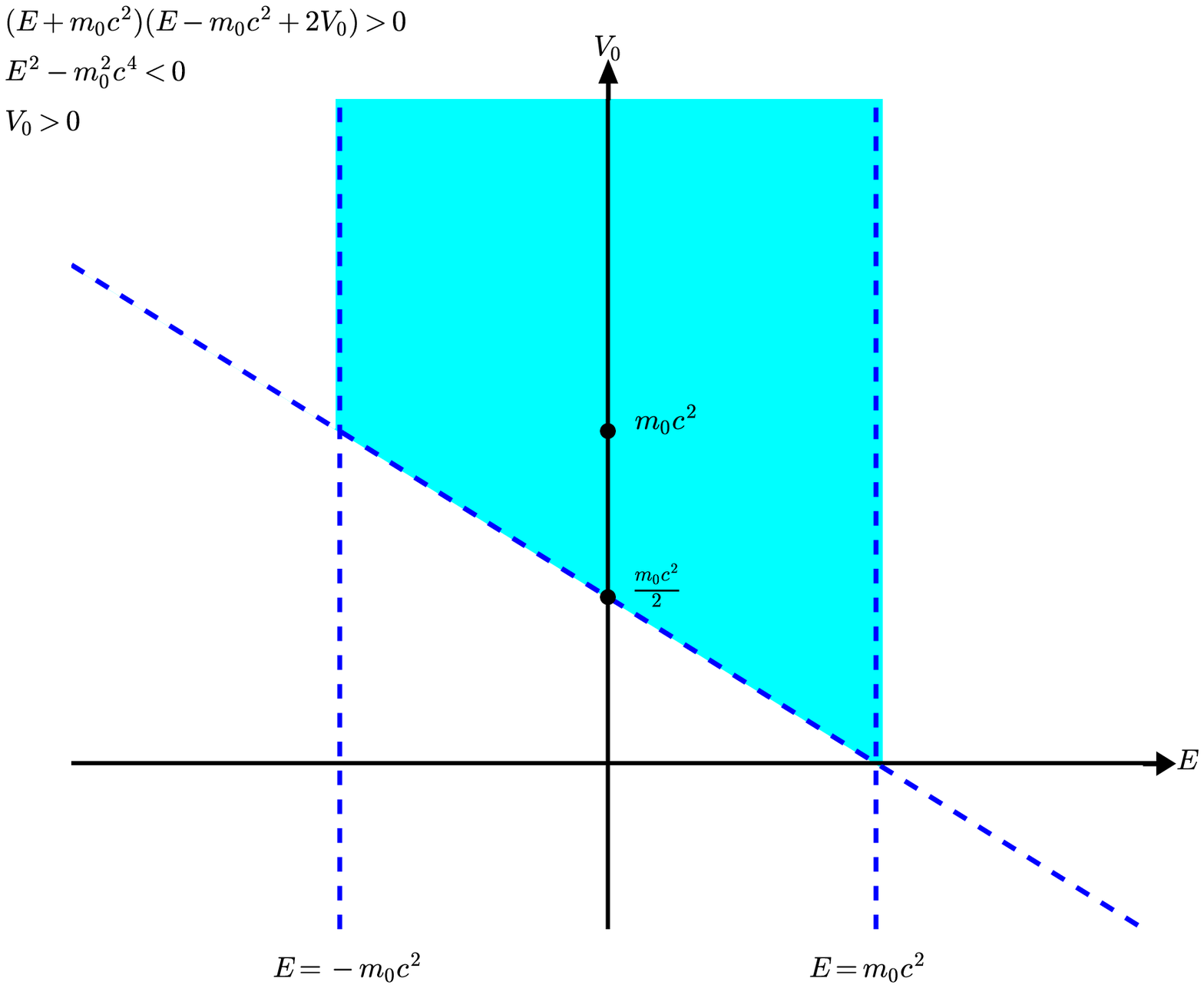}
   \caption{The results of the requirement to have bound states. The intersection area determines the correlation of the depth parameter with KG gap. } \label{fig:KGSSboundstates}
\end{figure}

The GSWS  potential well in one dimension has $V(-x)=V(x)$  symmetry. Therefore to study the problem in one region leads to determine the whole results. Moreover, this symmetry lets the energy spectrum be obtained from two subsets, even, $E_n^{e}$, and odd, $E_n^{o}$, spectra.
In $x<0$ region it is started by
  \begin{eqnarray}
  \Bigg[\frac{d^2}{dx^2}+\alpha^2\bigg[-\epsilon^2+\frac{\beta^2}{1+e^{-\alpha(x+L)}}+\frac{\gamma^2}
  {\big(1+e^{-\alpha(x+L)}\big)^2}\bigg] \Bigg]\phi_L(x) &=& 0. \label{KG2x<0}
\end{eqnarray}
where the parameters are
\begin{eqnarray}\label{epsilonbetagamma}
  -\epsilon^2&\equiv&\frac{(E^2-m_0^2c^4)}{\alpha^2\hbar^2c^2}\\
  \beta^2 &\equiv& \frac{2(E+m_0c^2)(V_0-W)}{\alpha^2\hbar^2c^2}  \\
  \gamma^2 &\equiv& \frac{2(E+m_0c^2)W}{\alpha^2\hbar^2c^2}.
\end{eqnarray}
In the negative region
\begin{eqnarray}
  z &=& \Big[1+e^{-\alpha(x+L)}\Big]^{-1}
\end{eqnarray}
is used for mapping and is found that the KG equation yields to the hypergeometric differential equations. Afterwards the analytical solutions are obtained for negative region as
\begin{eqnarray}
  \phi_L(z) &=& D_1 z^\mu (z-1)^{\nu} \,\,\, {}_2F_1[a_L,b_L,c_L;z]\nonumber\\
  &+&D_2 z^{-\mu} (z-1)^{\nu} \,\,\, {}_2F_1[1+a_L-c_L,1+b_L-c_L,2-c_L;z], \label{phileft}
\end{eqnarray}
while for the positive region
\begin{eqnarray}
  \phi_R(y) &=& D_3 y^\mu (y-1)^{\nu} \,\,\, {}_2F_1[a_R,b_R,c_R;y]\nonumber\\
  &+&D_4 y^{-\mu} (y-1)^{\nu} \,\,\, {}_2F_1[1+a_R-c_R,1+b_R-c_R,2-c_R;y]. \label{phiright}
\end{eqnarray}
in terms of  $D_1$, $D_2$, $D_3$ and $D_4$ normalization constants and
\begin{eqnarray}
  \mu &\equiv& \mp \sqrt{\epsilon^2}, \\
        &=& \frac{1}{\alpha}\sqrt{-\frac{E^2-m_0^2c^4}{\hbar^2 c^2}},  \\
  \nu &\equiv& \mp \sqrt{\epsilon^2-\beta^2-\gamma^2} ,  \\
        &=& \frac{i}{\alpha}\sqrt{\frac{(E+m_0c^2)(E-m_0c^2+2V_0)}{ \hbar^2 c^2}} ,  \\
  \theta &\equiv& \frac{1}{2}\mp \sqrt{\frac{1}{4}-\gamma^2}\\
         &=&\frac{1}{2}\mp \sqrt{\frac{1}{4}-\frac{2(E+m_0c^2)W}{\alpha^2 \hbar^2 c^2 }}.
\end{eqnarray}
parameters. Note that  six coefficients are defined  \begin{eqnarray}
  a_L &=& a_R\equiv \mu+\theta+\nu, \\
  b_L &=& b_R\equiv 1+\mu-\theta+\nu, \\
  c_L &=& c_R\equiv 1+2\mu.
\end{eqnarray}
Positive and real parameters $k_n$ and $\kappa_n$ are defined by
\begin{eqnarray}
  k_n &\equiv& \sqrt{-\frac{E^2-m_0^2c^4}{\hbar^2 c^2}}, \\
  \kappa_n &\equiv& \sqrt{\frac{(E+m_0c^2)(E-m_0c^2+2V_0)}{ \hbar^2 c^2}} .
\end{eqnarray}
Since the KG particle is embedded in GSWS well, it should have energy
\begin{eqnarray}
  -m_0c^2 < E< m_0c^2
\end{eqnarray}
within the KG gap. These conditions are studied in Fig. \ref{fig:KGSSboundstates}. If $V_0$ is greater than $0$ but less than $\frac{m_0c^2}{2}$ the bound states can only occur in the positive region of the KG gap. For greater $V_0$ but less than $m_0c^2$ the KG gap widens to the negative region. If the depth parameter is bigger than $m_0c^2$, then bound states expand to whole KG gap. In literature, it is found that this analysis has not been taken into account \cite{bayrakSahin2015}.

Before continuing, the attention of the readers should be focused on the linear correlation between $W$ and $V_0$. If these two parameters are independent, the surface effect could not play a role on the results. Therefore $W$ is not a  fabricated parameter.

The wave function solutions at infinities should fade.  Thus $D_2$ and $D_4$ should be zero. The continuity conditions dictate the wave function and its derivative should be  equal to each other at $x=0$ boundary. At zero two mapping parameters tend to a nonzero constant $t_0$,
\begin{eqnarray}
 t_0 &\equiv& \Big[1+e^{-\alpha L}\Big]^{-1}.
\end{eqnarray}
From the first condition it is found that
\begin{eqnarray}
  (D_1-D_3)t_0^\mu (t_0-1)^\nu M_1 = 0,
\end{eqnarray}
and similarly, from the second condition
\begin{eqnarray}
 (D_1+D_3)t_0^\mu (t_0-1)^\nu \Bigg[\Bigg(\frac{\mu}{t_0}+\frac{\nu}{t_0-1}\Bigg)M_1 +\frac{(\mu+\theta+\nu)(1+\mu-\theta+\nu)}{1+2\mu}M_3 \Bigg] =0
\end{eqnarray}
is derived. Here
\begin{eqnarray}
  M_1 &\equiv& {}_2F_1[\mu+\theta+\nu,1+\mu-\theta+\nu,1+2\mu;t_0], \\
  M_3 &\equiv& {}_2F_1[1+\mu+\theta+\nu,2+\mu-\theta+\nu,1+2\mu;t_0].
\end{eqnarray}
Since $t_0\approx 1$, the hypergeometric functions should be carefully examined around zero. Hence the transformation \cite{gradshytenBook}
\begin{eqnarray}
  _2F_1(a,b,c;q) &=& \frac{\Gamma(c)\Gamma(c-a-b)}{\Gamma(c-a)\Gamma(c-b)} \,\,\, _2F_1(a,b, a+b-c+1;1-q)\nonumber \\
  &&+(1-q)^{c-a-b}\frac{\Gamma(c)\Gamma(a+b-c)}{\Gamma(a)\Gamma(b)}\,\,\, _2F_1(c-a,c-b, c-a-b+1;1-q)\,\,\,
\end{eqnarray}
should be employed. Then $M_1$ and $M_3$ can be rewritten as
\begin{eqnarray}
  M_1 &=& S_1N_1+(1-t_0)^{-2\nu}S_2N_2, \\
  M_3 &=& S_3N_3+(1-t_0)^{-2\nu}S_4N_4,.
\end{eqnarray}
where
\begin{eqnarray}
    S_1 &=& \frac{\Gamma(1+2\mu)\Gamma(-2\nu)}{\Gamma(1+\mu-\theta-\nu)\Gamma(\mu+\theta-\nu)}, \\
    S_2 &=& \frac{\Gamma(1+2\mu)\Gamma(2\nu)}{\Gamma(1+\mu-\theta+\nu)\Gamma(\mu+\theta+\nu)},   \\
    S_3 &=& \frac{\Gamma(2+2\mu)\Gamma(-1-2\nu)}{\Gamma(1+\mu-\theta-\nu)\Gamma(\mu+\theta-\nu)}, \\
    S_4 &=& \frac{\Gamma(2+2\mu)\Gamma(1+2\nu)}{\Gamma(2+\mu-\theta+\nu)\Gamma(1+\mu+\theta+\nu)}.
\end{eqnarray}
and
\begin{eqnarray}
  N_1 &=& {}_2F_1[\mu+\theta+\nu,1+\mu-\theta+\nu,1+2\nu;1-t_0], \\
  N_2 &=& {}_2F_1[1+\mu-\theta-\nu,1+\mu+\theta-\nu,1-\nu;1-t_0], \\
  N_3 &=& {}_2F_1[1+\mu+\theta+\nu,2+\mu-\theta+\nu,2+2\nu;1-t_0], \\
  N_4 &=& {}_2F_1[1+\mu-\theta-\nu,\mu+\theta-\nu,-2\nu;1-t_0].
\end{eqnarray}
Consequently a subset of the energy spectra, $E_n^{e}$, is obtained by $D_1=D_3$ and
\begin{eqnarray}
 (D_1+D_3)t_0^\mu (t_0-1)^\nu \Bigg[\Bigg(\frac{\mu}{t_0}+\frac{\nu}{t_0-1}\Bigg)M_1 +\frac{(\mu+\theta+\nu)(1+\mu-\theta+\nu)}{1+2\mu}M_3 \Bigg] =0 \label{evensolutions}
\end{eqnarray}
while the other subset, $E_n^{o}$, is calculated by $D_1=-D_3$ and
\begin{eqnarray}
   M_1 = 0,\label{oddsolutions}
\end{eqnarray}
since $t_0^\mu$ and $(t_0-1)^\nu$ is nonzero. Note that the energy spectra  can be obtained via transcendental Eq.~\ref{evensolutions} and Eq.~\ref{oddsolutions}  numerically.
\section{Thermodynamics of a System}\label{sec:thermosection}

Once  micro states energy spectrum of a physical system is calculated,  the partition function can be written with  \cite{reifbook}
\begin{eqnarray}
  Z(\beta) &=& \sum_{n} e^{-\beta E_n}, \label{partitionfunction}
\end{eqnarray}
to express the Helmholtz free energy as follows
\begin{eqnarray}
      F(T) &\equiv& -k_B T\ln Z(\beta).
    \end{eqnarray}
Note that  $\beta$ is defined by the Boltzman constant $k_B$ and temperature $T$
\begin{eqnarray}
  \beta &=& \frac{1}{k_B T}.
\end{eqnarray}
From the fundamental thermodynamic relations, the entropy of the physical system is calculated by,
  \begin{eqnarray}
      S(T) &=& -\frac{\partial}{\partial T}F(T).
    \end{eqnarray}
Beside entropy, the expectation value of the energy of the system, the internal energy $U(T)$ is derived by
\begin{eqnarray}
      U(T) &=& -\frac{\partial}{\partial \beta}\ln Z(\beta).
    \end{eqnarray}
Finally, the measure of the energy required to raise the temperature $1K$ per unit mass, more precisely the isochoric specific heat capacity $C_v(T)$,  is defined by
\begin{eqnarray}
      C_v(T) &\equiv& \frac{\partial}{\partial T}U(T).
    \end{eqnarray}

\section{Results and Discussions}\label{sec:application}
In this section, modification of the thermodynamic functions by the surface interaction effects is going to be discussed. To establish the partition function, a confined particle and  GSWS potential well parameters should be chosen. In this manuscript neutral Kaon particle, that has rest mass energy $497.648MeV$, is taken into account within a GSWS potential well expressed with the chosen  parameters as follows:  $\alpha=1fm^{-1}$, $L=6fm$, $V_0=1.5 m_0c^2$ and $W=4 m_0c^2$. Note that all the parameters are determined arbitrarily. The Newton-Raphson method is used to solve  Eq.~\ref{evensolutions} and Eq.~\ref{oddsolutions}.  In Table~\ref{tab:energyspectrum WS} and Table~\ref{tab:energyspectrum GSWS}  the calculated energy spectra are tabulated for WS and GSWS potential wells, respectively.

\begin{table}[!ht]
\caption{\label{tab:energyspectrum WS} The energy spectrum of the neutral Kaon particle embedded in the artificial WS potential well.}
\begin{tabular}{|c|c|c|c|c|c|c|c|c|c|c|c|}
\hline    
$n$ &$E_n(MeV)$ &      $n$ &$E_n(MeV)$&      $n$ &$E_n(MeV)$&  $n$ &$E_n(MeV)$&  $n$ &$E_n(MeV)$ \\\hline
$0$&$-495.211$ &      $6$&$-277.810$&        $12$&$-12.567$    &$18$&$223.205$  &$24$&$415.503$\\
\hline
$1$&$-478.229$ &      $7$&$-232.367$&        $13$&$29.120$     &$19$&$258.811$  &$25$&$441.097$\\
\hline
$2$&$-448.020$ &      $8$&$187.198$&         $14$&$69.898$     &$20$&$293.174$  &$26$&$463.762$\\
\hline
$3$&$-409.842$ &      $9$&$-142.488$&        $15$&$109.750$    &$21$&$326.197$  &$27$&$482.471$\\
\hline
$4$&$-367.506$ &      $10$&$-98.433$&        $16$&$148.611$    &$22$&$357.729$  &$28$&$495.095$\\
\hline
$5$&$-323.059$ &      $11$&$-55.091$&        $17$&$186.454$    &$23$&$387.588$      &$$&$$\\
\hline    
\end{tabular}
\end{table}

\begin{table}[!ht]
\caption{\label{tab:energyspectrum GSWS} The energy spectrum of the neutral Kaon particle confined in the artificial GSWS potential well.}
\begin{tabular}{|c|c|c|c|c|c|c|c|c|c|}
\hline    
$n$ &$E_n(MeV)$ & $n$ &$E_n(MeV)$&  $n$ &$E_n(MeV)$&    $n$ &$E_n(MeV)$&    $n$ &$E_n(MeV)$\\\hline
$0$&$-492.093$       &$4$&$-280.055$   &$8$&$-28.447$
      &$12$&$206.642$  & $16$&$424.562$  \\
\hline
$1$&$-457.914$   &$5$&$-215.990$        &$9$&$32.022$      &$13$&$262.714$  &$17$&$476.444$  \\
\hline
$2$&$-404.742$   &   $6$&$-152.580$      &$10$&$91.307$     &$14$&$317.693$ & $$&$$  \\
\hline
$3$&$-343.541$  &$7$&$-89.941$  &$11$&$149.544$  &$15$&$371.663$ &$$&$$
\\
\hline    
\end{tabular}
\end{table}

\begin{figure}[!htb]
\centering
\includegraphics[totalheight=0.5\textheight]{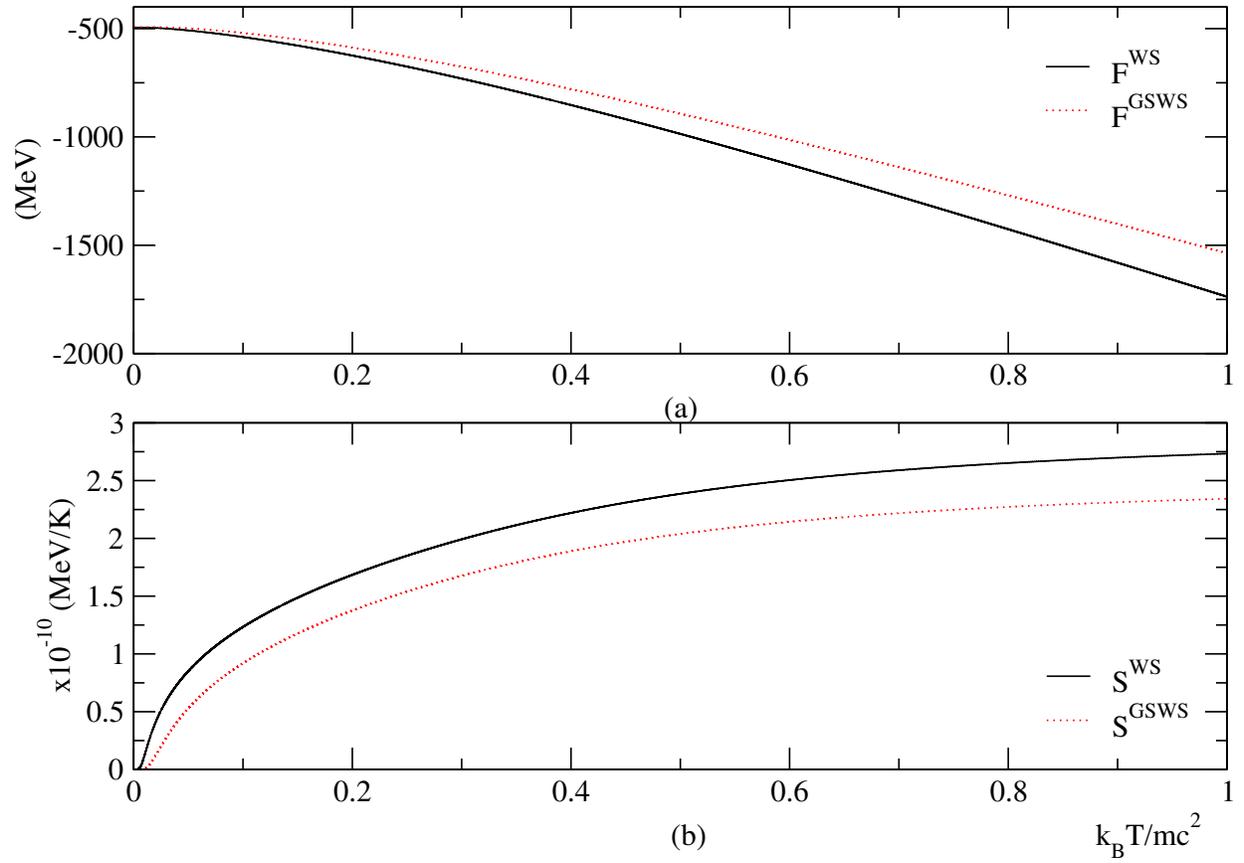}
\caption{Helmholtz energy $F(T)$ (a), entropy $S(T)$ (b), as functions of reduced temperature.}
\label{fig:KGHelmholtzFreeEnergyandEntropyWSvsGWSW}
\end{figure}

Using Equation \ref{partitionfunction}  the partition function is constituted. The plots of the Helmholtz free energy and the entropy functions versus reduced temperature are given in Figure \ref{fig:KGHelmholtzFreeEnergyandEntropyWSvsGWSW} (a) and (b), respectively. In the relativistic case similar to non-relativistic one, the entropy starts from zero at $0K$  and complies with the third law of thermodynamics.
As expected, the number of the existing micro states decreases by the surface interaction effects
accompanied by the upward shift in the spectrum and results in the increase in the Helmholtz free energy function.

\begin{figure}[!htb]
\centering
\includegraphics[totalheight=0.5\textheight]{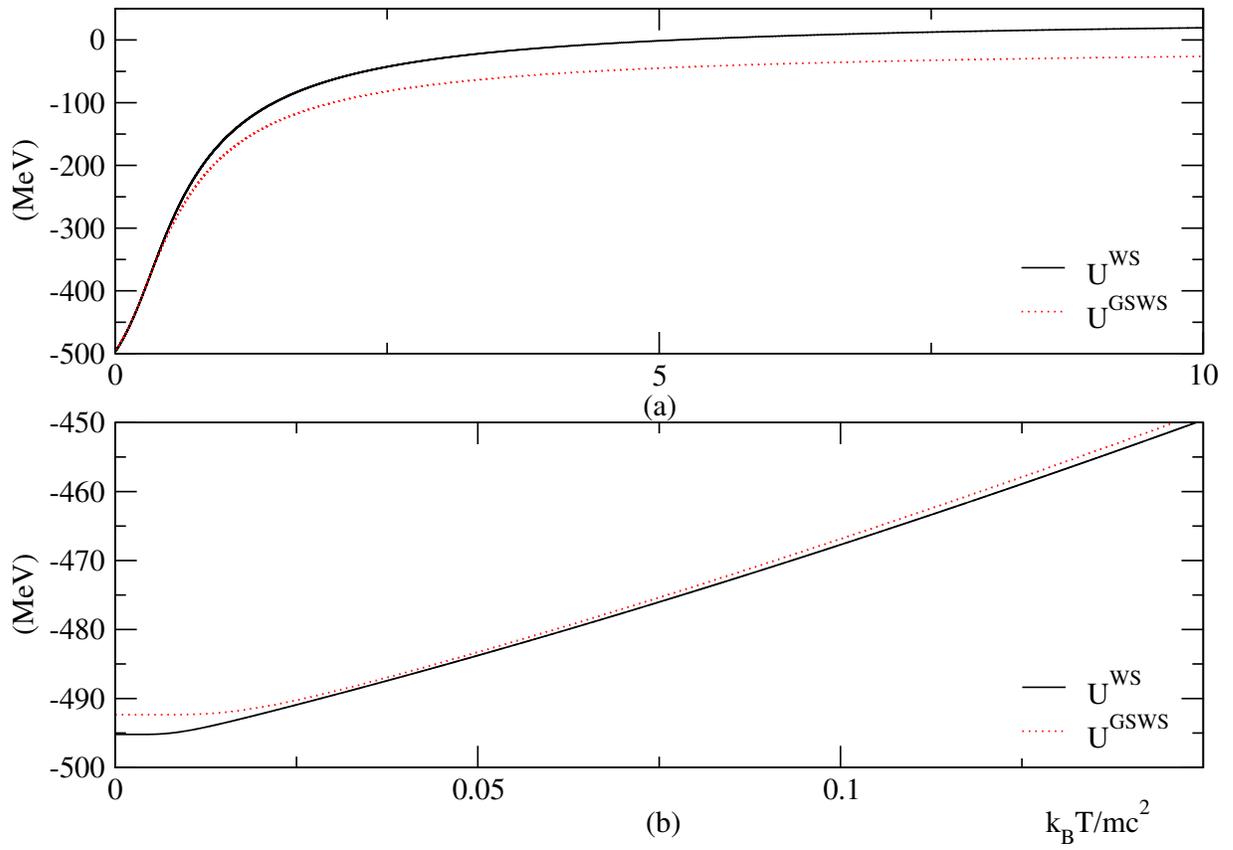}
\caption{Internal energy $U(T)$ as functions of reduced temperature (a), the initial behavior (b).}
\label{fig:KGKaonInternalEnergyWSvsGWSW}
\end{figure}

Afterwards the internal energy $U(T)$ function is investigated  in Figure~\ref{fig:KGKaonInternalEnergyWSvsGWSW} (a). The results differ from the non-relativistic problem. The expectation value of the energy saturates to a positive value in WS potential well while results in a negative value in GSWS well. They are calculated as $40.022MeV$ and $-7.387MeV$, respectively. The initial behavior of internal energy versus the reduced temperature is given in Figure~\ref{fig:KGKaonInternalEnergyWSvsGWSW} (b). At $0K$ the initial internal energies are consistent with microstate's energy spectra. Similar to the non-relativistic case, indistinguishable internal energies become distinguishable when the reduced temperature increases.

\begin{figure}[!htb]
\centering
\includegraphics[totalheight=0.5\textheight]{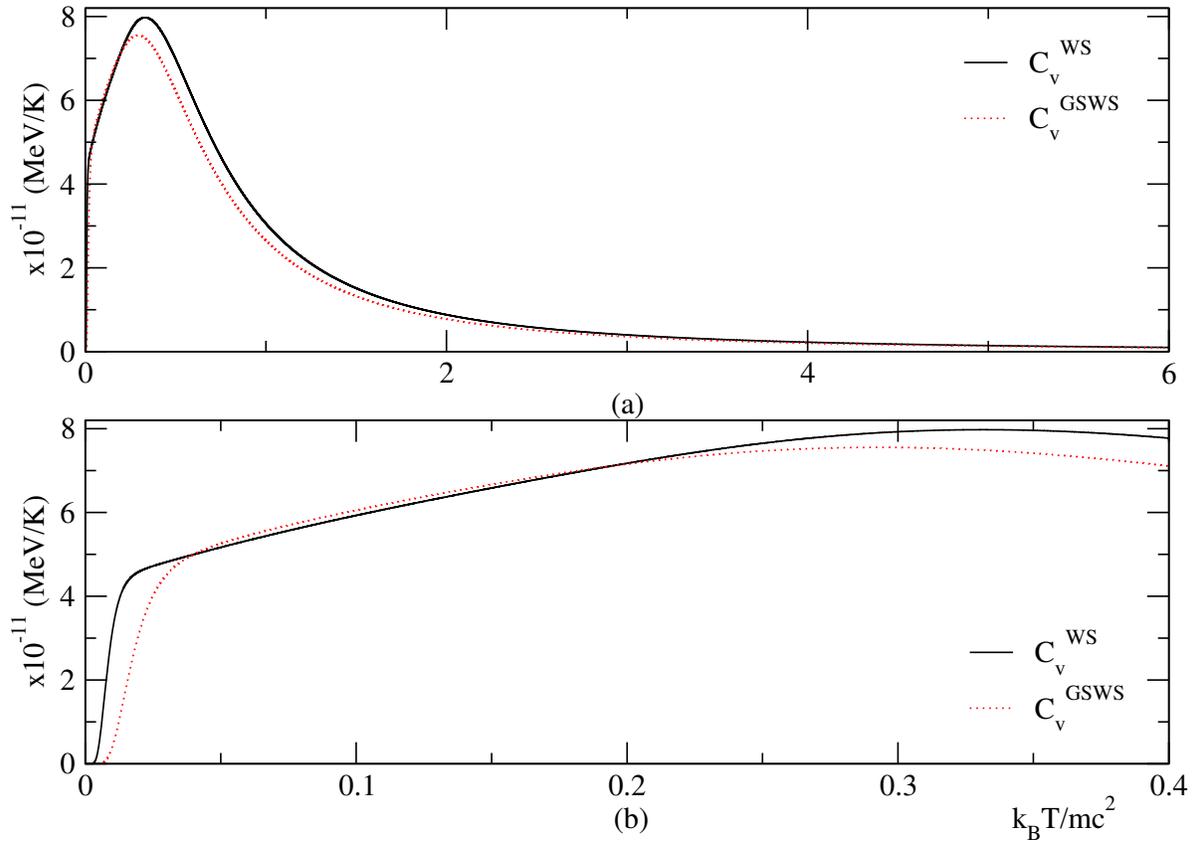}
\caption{Specific heat $C_v(T)$ as functions of reduced temperature (a), the initial behavior (b).}
\label{fig:KGKaonSpecificHeatandInsetWSvsGWSW}
\end{figure}

Finally in Figure~\ref{fig:KGKaonSpecificHeatandInsetWSvsGWSW} (a), the specific heat $C_v(T)$ versus reduced temperature is investigated. In contrast with the non-relativistic problem, it is found that in WS well the neutral Kaon reaches a higher value of specific heat at a higher reduced temperature compared with GSWS well. The demonstration of the initial behavior of the specific heat is plotted in Figure~\ref{fig:KGKaonSpecificHeatandInsetWSvsGWSW} (b). Note that it points out that at least two values of reduced temperature either in WS or GSWS well becomes indistinguishable for the confined neutral Kaon particle.

Before the conclusion section two more points can be clarified:

The first one is: "Is it possible to obtain analytically closed formulas of the thermodynamic functions within this problem?". Unfortunately, the answer is "no", because the energy spectrum is obtained by solving transcendental equations expressed in Eq.~\ref{evensolutions} and Eq.~\ref{oddsolutions}.

The second one is, how the corresponding equations could be modified if the three-dimensional solutions are done. Note that in such problem, the angular momentum term takes place in the effective potential. Usually, the exact solution of the KG equation in most potentials like GWS for the $l\neq 0$ states cannot be obtained in three dimensions.  To deal with the orbital term, famous approximations such as Pekeris \cite{pekeris1934},  Green and Aldrich \cite{greenetal1976} can be used and already examined by \cite{bayrakAciksoz2015}.  For $l=0$ states, the radial equation which has different boundary conditions should be solved. In that case, a different energy spectrum will be obtained \cite{bayrakSahin2015}, though the thermodynamic functions might vary.

\section{Conclusion}\label{sec:conclusion}

In this manuscript,  the bound state solution of Klein-Gordon equation with mixed scalar-vector generalized symmetric Woods-Saxon potential is studied. The required conditions are obtained in order to confine a particle in the potential energy well. Moreover, the energy spectrum equations are found in the transcendental form. As an application, to calculate the energy spectrum,   the neutral Kaon particle is chosen. The potential well is determined with arbitrary parameters excluding and including the surface effects. By using the partition function the thermodynamic functions versus reduced temperature are investigated. Since the energy spectrum shifted by the additional terms discussed, it is observed that the Helmholtz free energy shifts similar to the non-relativistic case. Moreover, the decrease of the available number of energy eigenvalue caused a decrease in entropy, as expected. It is found out that the relativistic case differs from the non-relativistic case in terms of internal energy and specific heat functions. Without surface effects, internal energy saturated with a positive value, while with effects are considered with a negative value. As a consequence, the specific heat function sailed to a higher value at a higher reduced temperature in Woods-Saxon well.

\section*{Acknowledgements}
The author would like to thank Dr. M. Erdogan, Dr. E. Pehlivan, C. Ertugay and finally to an anonymous referee for his/her thorough review and highly appreciate the comments and suggestions, which significantly contributed to improving the quality of the publication.

\newpage

\newpage

\newpage

\newpage

\end{document}